\documentclass[aps,twocolumn,showpacs,showkeys]{revtex4}
\usepackage{psfig}
\usepackage{graphicx}
\usepackage{amssymb}
\usepackage{bm}
\begin{document}
\newcommand{\etal}{{ et al. }} 
\newcommand{\sy}{Seyfert }
\newcommand{\be}{\begin{equation}}
\newcommand{\ee}{\end{equation}}
\newcommand{\msun}{{\rm M_{\odot }}}
\setcounter{page}{1}
\title[]{Black hole mass and accretion rate of Active Galactic Nuclei}
\author{Xue-Bing \surname{Wu}, F.K. Liu, R. Wang}
\affiliation{Department of Astronomy, Peking University, Beijing
100871, China}
\author{M.Z. \surname{Kong}, J.L. Han}
\affiliation{National Astronomical Observatories, Chinese Academy of Sciences,
Beijing 100012, China}
\date[]{Received February 28 2006}

\begin{abstract}
The determination of the central black hole mass is crucial to the
understanding of AGN physics. In this paper we briefly review some
methods that are currently used to
estimate the black hole mass of AGNs. Particularly we demonstrate the 
importance of two correlations: one
 between the black hole mass and the stellar velocity
dispersion and another one between the broad line region size and the
optical continuum luminosity. Besides applying these relations in deriving 
black hole masses of various types of AGNs, we also employed the fundamental plane
of elliptical galaxies to estimate the central velocity dispersions of AGN host
galaxies, and then the black hole masses of AGNs including BL Lac objects.
In addition, we derived another empirical relation between the BLR size 
and H$_\beta$ emission line luminosity from  AGNs with the BLR size
measured by reverberation mapping studies, and 
argued that more accurate black hole masses of extremely 
radio-loud AGNs could be obtained with it than using the 
usual $R-L_{5100\AA}$ relation
because of the jet contribution to the optical continuum. Finally
we pointed out that black hole mass estimation is very much helpful to 
determine the accretion rate and understand the accretion process in AGNs.
\end{abstract}



\maketitle

\section{INTRODUCTION}
As we know, mass is the most important physics parameter of a black hole. Many
other properties of the black hole are directly related to its mass. 
Currently astrophysical black holes can be divided into three categories: the 
primordial black holes with the mass of about $10^{15}g$ existed only in the early 
universe, the stellar-mass black holes with the mass of about 10 solar masses
($M_\odot$) existed in some X-ray binaries and the supermassive black holes 
with the mass of $10^6$ to $10^9M_\odot$ existed in the center of normal and
active galaxies. In addition, recently there were also some indications of the possible
existence of
the intermediate-mass black holes (with the mass of $10^2$ to  
$10^4 M_\odot$) in some ultra-luminous X-ray sources, but this is still 
under debate. Black holes were also suggested to exist in gamma-ray bursts and a 
magnetized accretion disk around the black hole may contribute significantly
to the energy output of them (Lee \& Kim 2002; Kim \& Lee 2003). 
In this paper we mainly focus on the supermassive black holes in
Active Galactic Nuclei (AGNs), which consist of many powerful extragalactic
sources including quasars, radio galaxies, Seyfert galaxies, BL Lacertae 
objects, LINERs, etc. Supermassive black holes (hereafter SMBHs) 
 have been suggested to exist in the center
of AGNs in order to account for the huge power
 of these energetic extragalactic objects (Lynden-Bell 1969; Rees 1984).

In the last two decades, SMBHs have been found in the 
center of about 40 nearby galaxies (including our Milky Way) as well, 
especially with the Hubble Space 
Telescope and other giant ground-based telescopes (See Kormendy \& Gebhardt
2001 for a recent review). These discoveries were made by three dynamical methods, 
involving the studies on the dynamics of stars, gas and extragalactic water 
masers respectively. However, these direct methods for black hole mass 
measurements can not be applied to most of AGNs because (1) AGNs are so bright 
to outshine the stars, and (2) only in several nearby AGNs we can observed
the circular motions of their central gas and water masers. Previously, some
indirect methods, such as fitting the observed UV/topical 'Big Blue Bump' and 
the X-ray ion $K_\alpha$ line in the spectra
of AGNs using the accretion disk models, have been adopted to estimate the 
central black hole mass of AGNs. But these estimates are highly uncertain 
because they depends on our knowledge of the disk inclination, accretion rate
 and radiation emissivity law, and all of them are poorly known.
 Probably the most reliable current method for AGN black hole mass estimation 
is the reverberation mapping. Using this technique, the size
of the broad line region (BLR) can be measured using the time lag between the
variabilities of continuum and emission line fluxes. The black hole mass can be
then estimated from the BLR size and the characteristic velocity (determined
by the full width at half-maximum (FWHM)
of the emission line) using the virial dynamics. So far, reverberation mapping
studies have
yielded black hole masses of about 20
Seyfert 1 galaxies and 17 nearby bright quasars (Wandel \etal 1999;
Ho 1999;
Kaspi \etal 2000; Peterson \etal 2004). However, even for the reverberation mapping
study, there are also some uncertainties of the results mainly due to the
unknown BLR dynamics and geometry (Krolik 2001; Wu \& Han 2001a).

With the estimated black hole masses, a tight 
correlation between black hole mass and bulge velocity
dispersion ($\sigma$) has been found for nearby galaxies 
(Tremaine et al. 2002) and for a few Seyfert galaxies as well (Ferrarese et al.
 2001).  This implies that the  M$_{BH}$-$\sigma$ relation is probably  
universal for both active and inactive galaxies. Such a  
tight relation suggests an interesting
possibility to estimate the central black hole masses for galaxies using the
measured values of bulge velocity dispersions. This straightforward
method is particularly important for  AGNs. Especially BL Lacertae objects,
 the reverberation mapping technique cannot be applied  because they 
have no or only very weak emission lines in their optical spectra. Using
the   M$_{BH}$-$\sigma$ relation may be the only possible way to derive
the black hole masses for them.

An empirical relation between the 
BLR size ($R$) and the optical continuum luminosity at 5100$\AA$ ($L_{5100\AA}$) 
has been derived by Kaspi et al. (2000) 
using the observed data of 34 nearby AGNs in the reverberation mapping studies. 
This relation has been frequently adopted to estimate the BLR size and then
derive the black hole masses for large samples of AGNs, including radio-loud 
quasars (Gu, Cao \& Jiang 2001; Oshlack, Webster \& Whiting 2001). 
However, the optical luminosity of  some radio-loud AGNs may not
be a good indicator of ionizing luminosity, which is usually related to the 
UV/optical radiation from the accretion disk around the central black hole. 
The powerful jets of blazar-type AGNs may significantly contribute to the 
optical luminosity. This has been confirmed by the discovery of optical jets
 in AGNs by HST (Scarpa et al. 1999), and the detected optical synchrotron 
radiation in many other radio-loud AGNs (Whiting, Webster \& Francis 2001). 
Therefore, using the  $R-L_{5100\AA}$ empirical relation,  
which was obtained based on the sample of 
mostly radio-quiet AGNs, one would significantly overestimate the actual
BLR size and hence the black hole mass of radio-loud AGNs. 
Oshlack et al. (2002) have shown that their estimated black hole
masses would be lower if the synchrotron contribution to the optical flux is 
subtracted.

In this paper, we will present some results in using the empirical relations 
( M$_{BH}$-$\sigma$ relation and  $R-L_{5100\AA}$ relation) 
to estimate the SMBH masses of different types of AGNs. We will also
report some other progress, including using the fundamental plane relation
of the elliptical host galaxies and  
a new $R-L_{H_\beta}$ empirical relation  to estimate the black hole
masses of radio-loud AGNs. In addition, based on the obtained black hole mass 
we will study the accretion rate of some AGNs and discuss the accretion 
physics around black holes.  

\section{Applications of the $\rm M_{\rm BH}$-$\sigma$ relation}
\subsection{Black hole masses of Seyfert galaxies and their relation with 
bulge masses}
Nelson \& Whittle (1995) have compiled the
measurements of nuclear
velocity dispersions of about 70 Seyfert galaxies. After excluding some LINERs
and normal galaxies, we got a sample of 
33 Seyfert 1s and 32
Seyfert 2s (Wu \& Han 2001b). We estimated the SMBH masses of these Seyfert galaxies directly
using the 
$\rm M_{\rm BH}$-$\sigma$ relation given by Gebhardt et al. (2000).
The V-band bulge absolute magnitude was estimated from the B-band 
 bulge absolute magnitude  in Nelson \& Whittle (1995) by assuming
the B-V color of the bulge being the same as the total B-V color of Seyfert 
galaxy.

\begin{figure}
\psfig{file=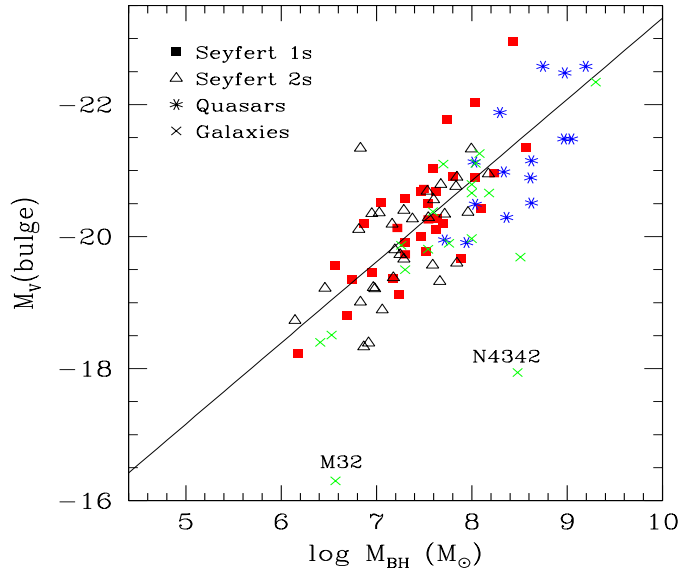,width=9.5cm,height=8cm,angle=0}
\caption{Correlation of V-band absolute bulge magnitudes and SMBH masses 
for quasars, normal galaxies and Seyfert galaxies. The solid line
represents the least-squares fits to all objects. Figure taken from Wu \& Han
2001b.}
\end{figure}

The relation between V-band bulge luminosities and SMBH masses for 33 Seyfert 
1s and 32 Seyfert 2s is shown in Figure 1. 
The black hole masses of Seyfert galaxies are mostly in the range of $10^6$ 
to $10^8M_\odot$.
In Figure 1 we can clearly observe a significant universal
relation between the V-band bulge luminosities and SMBH masses for both Seyfert 
galaxies and quasars, as well as for nearby galaxies. For comparison, the SMBH 
masses of 15 quasars and the absolute 
V magnitudes of their inner host were taken from Laor (2001).
The SMBH masses determined by
stellar dynamics for 16 nearby galaxies were taken from G00
and their  absolute bulge
V magnitudes were taken from Laor (2001). Adopting the `typical' uncertainties 
of $M_V^{bulge}$  and 
$\rm M_{\rm BH}$, the least-squares fit for all objects gives,
\begin{equation}
M_V^{bulge}=-12.14\pm0.48-(1.08\pm0.06)\rm{log}(\rm M_{\rm BH}/M_\odot),
\end{equation}
The Spearman rank-order correlation coefficient is $r_S=-0.74$, 
which has a probability of 
$P_r = 1.3 \times 10^{-17}$ to occur by chance.   
Adopting the standard relation 
$ M_V^{bulge} = 4.83 -2.5 {\rm{log}} {L_{bulge}/L_\odot}, $
and the relation between the bulge mass and luminosity (Magorrian \etal 1998), 
from eq. (2) we get, 
${\rm M_{\rm BH} \propto M}_{bulge}^{1.96 \pm 0.13}$. 
Laor (2001) found ${\rm M_{\rm BH} \propto M}_{bulge}^{1.54 \pm 0.15}$ for 
a sample of objects including 9 \sy galaxies and 15 quasars.
Our result shows that this nonlinear relation is more evident from a larger 
sample including more \sy galaxies. 
Using another well-selected sample including 22 Seyfert 1s and 15
Seyfert 2s (Wu \& Han 2001b), we also found $M_{BH}\propto M_{bulge}^{(1.74\pm 0.14)}$.
Therefore, the nonlinear 
$\rm{M_{BH}}$-$\rm{M_{bulge}}$ relation is confirmed by our studies.

\subsection{Fundamental plane relation and black hole masses of BL Lac objects, 
radio galaxies and quasars}

AGNs usually have very bright nuclear emission, which makes it difficult
to measure their stellar velocity dispersions with the spectroscopic method.
 It is well known for  
ellipticals that three observables, namely the effective radius, 
the corresponding average surface
brightness and the central velocity dispersion, follow a surprisingly
tight relation (so-called the fundamental plane). 
Some studies have shown that  the elliptical hosts of radio 
galaxies follow the same fundamental plane
as normal ellipticals (Bettoni et al. 2001).
For about 300 normal ellipticals and radio galaxies,
Bettoni et al. (2001) found that the fundamental plane can be robustly 
described as 
$$
\log R_e = (1.27\pm0.04) \log \sigma + (0.326\pm0.007) <\mu_e>_R
$$
\begin{equation}
~~~~~~~~~~~~~~~~~~~~~~~~~~~~~~~~~~~~~~~~- 8.56\pm0.06,
\end{equation}
where $R_e$ is the effective radius in $kpc$, $\sigma$ is the central velocity
dispersion in $km s^{-1}$, and $<\mu_e>_R$ is the average surface brightness 
in R-band. Because the fundamental plane probably exists also for the
elliptical hosts of AGNs, this provides
a possible way to estimate the central velocity dispersions and then
the SMBH masses of AGNs.

\begin{figure}
\psfig{file=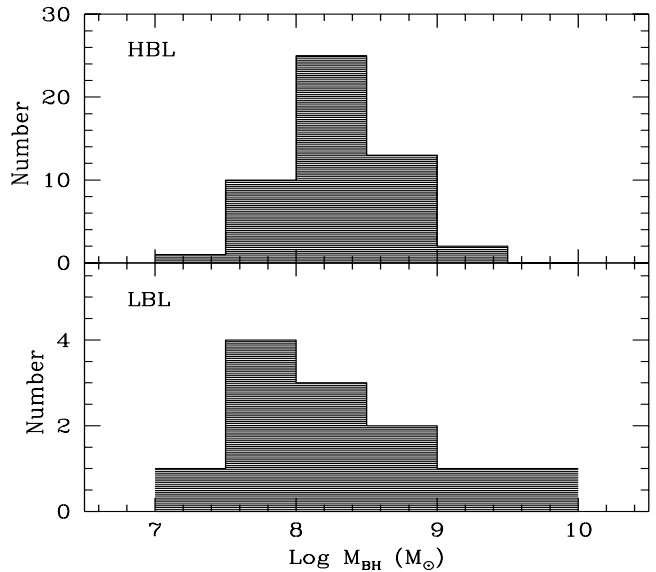,width=8.5cm,height=7.5cm,angle=0}
\caption{Histogram of the derived SMBH mass distribution of HBLs and LBLs 
using the fundamental plane and the  M$_{BH}$-$\sigma$ relation. 
Figure taken from Wu \etal (2002). }
\end{figure}

The BL Lac snapshot survey using the HST has obtained images for 110
BL Lac objects (Scarpa et al. 1999).  Urry et al. 
(2000) has obtained
host galaxy parameters for 72 of these BL lacs.  Among them, 63 objects
have measured redshifts. 51 of them are classified as high-frequency 
peaked BL Lacs (HBL)
and 12 of them as  low-frequency peaked BL Lacs (LBL).  Figure 2  shows the
histograms of our derived black hole masses based on the fundamental plane and 
the  M$_{BH}$-$\sigma$ relation for HBLs and LBLs (Wu, Liu \& Zhang 2002).
It is clear that there is no significant difference in
the black hole masses between HBLs and LBLs. 
 Most BL Lacs have black hole masses in the
range of $10^{7.5}M_\odot$ to  $10^{9.5}M_\odot$. 

\begin{figure}
\psfig{file=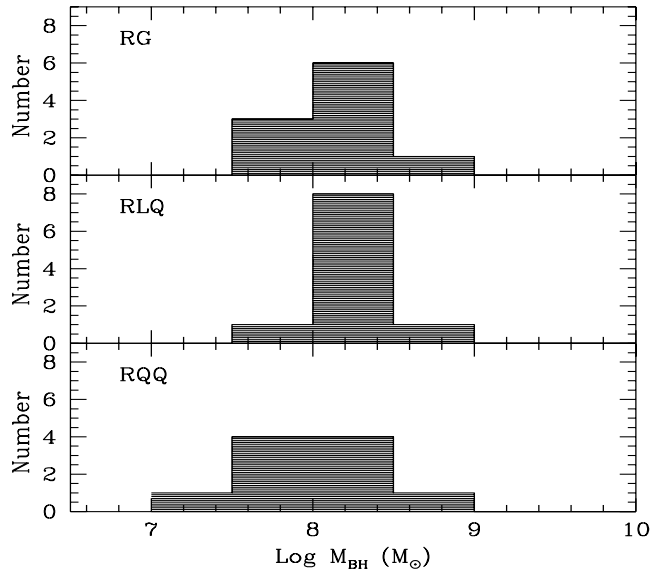,width=8.5cm,height=7.5cm,angle=0}
\caption{Histogram of the derived SMBH mass distribution of radio galaxies, 
radio-loud and radio-quiet quasars using the fundamental plane and the  
M$_{BH}$-$\sigma$ relation. Figure taken from Wu \etal (2002).}
\end{figure} 

A deep HST imaging study of the host galaxies of a sample of 10 radio 
galaxies (RGs), 10 radio-loud
quasars (RLQs) and 13 radio-quiet quasars (RQQs) has been  performed 
(McLure et al. 1999).  The hosts of both radio-loud AGNs and bright 
radio-quiet AGNs were found to be virtually all massive ellipticals. Therefore,
 we can also derive the central
velocity dispersions and black hole masses for these AGNs based on the 
morphology parameters of their host galaxies (Wu \etal 2002).
Figure 3 shows  the
histograms of the derived black hole masses for these AGNs.
It is clear that there are no significant differences in the 
SMBH masses among RGs, RLQs and RQQs. 
Most of these AGNs have black hole masses in the
range of $10^{7.5}M_\odot$ to  $10^{9}M_\odot$. Our results indicate that 
there is no
difference in SMBH masses of BL Lacs, RGs and RLQs. 
Different from some previous claims that
RLQs have more massive SMBHs than RQQs
(Laor 2000), our result shows that there is only a weak  difference in our 
derived SMBH masses for RQQs and RLQs in this sample if their host galaxies
are ellipticals.

Using the same method, we also estimate the primary black hole mass of another
well-known BL Lac object OJ 287 (Liu \& Wu 2002). Our result gives the mass 
about
$4\times10^8M_\odot$, which is consistent with the upper limit ($10^9M_\odot$)
obtained by Valtaoja
et al. (2000) based on a new binary
black hole model for OJ 287.

\section{Application of the $R-L_{5100\AA}$ 
relation}

Reverberation mapping studies of 34 AGNs revealed a significant correlation 
between
the broad line region (BLR) size and the optical continuum luminosity 
(Kaspi et al. 2000), namely, 
\be
R_{BLR}=32.9[\frac{\lambda L_{5100\AA}}{10^{44}erg/s}]^{0.7}~ \rm{lt-days} .
\ee
This empirical relation provides a possible way to estimate the black hole
mass of an AGN from a single optical spectrum, from which 
the continuum luminosity can be easily measured. On the other hand, 
the width (FWHM) of Balmer lines, which  can be
used to estimate the velocity dispersion of the BLR. Therefore, The virial
mass of the central black hole can be estimated with
\be
M_{BH}=1.464\times10^5(\frac{R_{BLR}}{\rm{lt-days}})(\frac{V_{FWHM}}
{10^3km/s})^2 \msun ,
\ee
where $V_{FWHM}$ is the FWHM value of Balmer emission lines. 

Here we show some of our results in estimating the black hole mass of double-peaked
broad line AGNs using the $R-L_{5100\AA}$ relation (Wu \& Liu 2004). These AGNs
represent only a small fraction of AGNs and the double-peaked
emission lines are usually attributed to the line emission from  
accretion disks. So far about 150 double-peaked broad emission line AGNs have been 
discovered. 
In a completed survey, Eracleous \& Halpern (1994, 2003) have found
26 radio-loud double-peaked broad emission line AGNs. Recently Strateva et al. (2003) 
presented a new sample of 116 double-peaked Balmer line AGNs selected from
the Sloan Digital Sky Survey (SDSS). 
In addition, double-peaked broad emission lines
have been also found in some nearby galaxies, including NGC 1097,  M81, 
NGC 4450,
and NGC 4203. These objects are all classified 
spectroscopically as Type 1 LINERs with very smaller nuclear luminosity
($<10^{43}erg/s$) and very lower Eddington ratios ($<10^{-3}$)(Ho et al. 2000).

 Strateva et al. (2003) have
given the magnitudes (Galactic extinction corrected), redshifts and FWHM values of
the double-peaked H$_\alpha$ line for 109 SDSS AGNs (7 objects have no FWHM
values among 116 AGNs in their table 3). From the $g$ and $r$ magnitudes of these AGNs
(available in Tables 1 and 2 in Strateva et al. (2003)), we can
estimate the rest frame luminosity at 5100$\AA$. Using the conversion formula between 
the FWHM values of H$_\alpha$
and H$_\alpha$ lines,$V_{FWHM}({\rm{H_\alpha}})=0.873V_{FWHM}
({\rm{H_\beta}})$, we can estimate the black hole masses of the double-peaked
SDSS AGNs with Eqs (3) and (4). The same method can be applied to 26 double-peaked 
AGNs found from a survey of
 radio-loud emission line 
AGNs (Eracleous \& Halpern 1994, 2003). The Eddington ratios ($L_{bol}/L_{Edd}$, where
$L_{bol}$ is the bolometric luminosity and $L_{Edd}$ is the Eddington luminosity) 
of these AGNs can be also
derived if we adopted $L_{bol}\simeq 9L_{5100\AA}$ (Kaspi \etal 2000). 
For  109 SDSS double-peaked AGNs, the black hole masses of are in the range from 
$4\times10^7 \msun$ to $5\times10^9\msun$, and the Eddington ratios are between 
0.002 and 0.2. The average black hole mass of these AGNs is $10^{8.74}M_\odot$
and the average Eddington ratio is $10^{-1.82}$. 
For the 26 radio-loud double-peaked AGNs, the black hole masses are from $3\times10^7
\msun$ to $3\times10^9\msun$ and the Eddington ratios are between 0.001 and 0.08,
The average black hole mass of these 26 double-peaked AGNs is $10^{8.78}M_\odot$
and the average Eddington ratio is $10^{-2.01}$, both being similar as those obtained 
for the double-peaked SDSS AGNs. Figure 4 summarized our results. Clearly, our derived
Eddington ratios for the SDSS and radio-loud double-peaked AGNs are larger
than the LINER-type double-peaked sources, which may reflect the difference in the 
accretion process for these AGNs (see section V for discussions).

\begin{figure}
\psfig{file=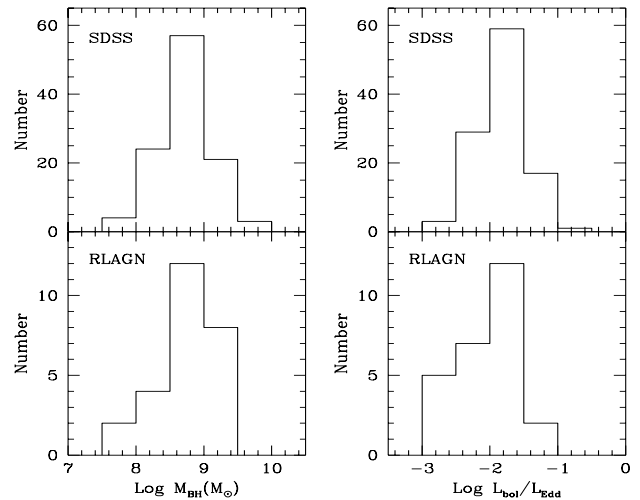,width=9cm,height=7.5cm,angle=0}
 \caption{The histograms of the black hole masses and Eddington ratios of 
double-peaked AGNs in the SDSS and radio-loud AGN samples. Figure taken from
Wu \& Liu (2004).
} 
\end{figure}

\section{Black hole masses derived from the  BLR size and H$_\beta$ luminosity 
relation}

\subsection{The $R-L_{H_\beta}$ relation}

\begin{figure}
\centering
\psfig{file=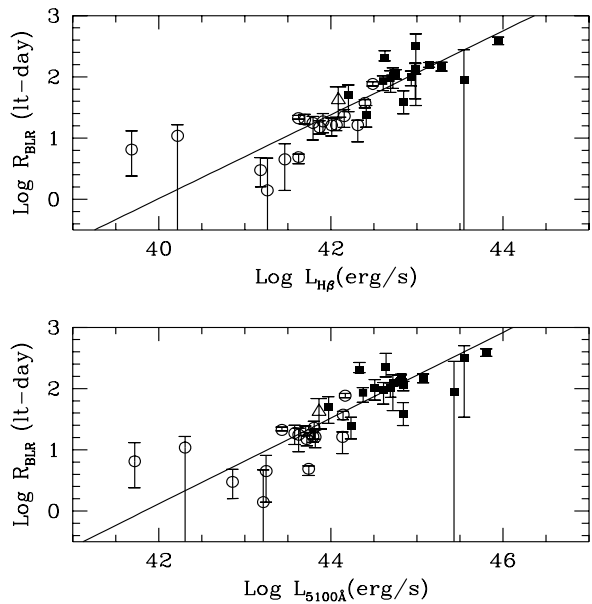,width=9cm,height=8.5cm,angle=0}
\caption{ The upper panel shows
the $R-L_{H_\beta}$ relation. The line shows the OLS bisector fit to the data.
 The lower panel shows the $R-L_{5100\AA}$ 
relation, which is 
identical to that in Kaspi \etal (2000) except that we include Mrk 279 (open triangle). 
The line
represents the linear fit given in Kaspi \etal (2000). The open and filled symbols 
denote Seyferts and
quasars respectively. Figure taken from Wu \etal (2004).}
\end{figure}

Using the available data of BLR sizes 
and H$_\beta$ fluxes for 34 AGNs in the reverberation mapping studies, we can 
investigate the relation between the BLR size and the
H$_\beta$ emission line luminosity (Wu \etal 2004). 
The H$_\beta$ flux data were available for  16 PG quasars and 17 
Seyfert 1 galaxies.
Because there is no available data of H$_\beta$ flux
for PG 1351+640 (Kaspi \etal 2000), we exclude this object from our investigation.
In addition, we add another Seyfert galaxy Mrk 279  in our sample because both the BLR
size and the H$_\beta$ flux have been measured recently (Santos-Lleo \etal 2001). 
All H$_\beta$ luminosity data  have been corrected for  
Galactic extinction using the values from NED. With these data, we derive an empirical
relation between the BLR size 
and H$_\beta$ luminosity:
$$
\rm{Log}~R~(\rm{lt-days}) = (1.381\pm0.080)+~~~~~
$$
\be
~~~~~(0.684\pm0.106) Log~(L_{H_\beta}/10^{42}~ergs~s^{-1}) .
\ee
The Spearman's rank correlation coefficient of our $R-L_{H_\beta}$ relation is 0.91, 
slightly higher than 0.83 for the $R-L_{5100\AA}$ relation (Kaspi \etal 2000).

In Fig. 5 we show the dependence of the BLR size on $L_{H_\beta}$ and $L_{5100\AA}$. 
Obviously these two relations are similar, which 
means that the $R-L_{H_{\beta}}$ relation can 
be an alternative of the  $R-L_{5100\AA}$ relation in estimating the BLR size for 
radio-quiet AGNs.

\subsection{Comparison of black hole mass estimation of AGNs from two R-L relations}
Since the $R-L_{5100\AA}$ relation obtained by Kaspi \etal (2000) has been frequently 
used to estimate the BLR size and the black hole mass of both radio-quiet and 
radio-loud
 AGNs, it is important to investigate the applicability of such an approach 
for radio-loud objects. 

\begin{figure}
\centering
\psfig{file=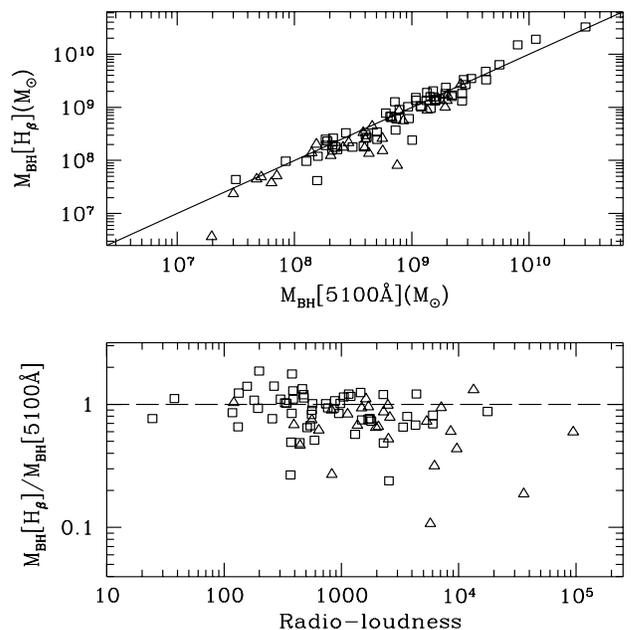,width=9cm,height=8.5cm,angle=0}
 \caption{Upper panel: Comparison of the black hole masses of radio-loud quasars 
estimated with $R-L_{H_\beta}$ and 
$R-L_{5100\AA}$ relations. The squares represent 59 quasars in Brotherton (1996)
and the triangles represent 27 PHFS quasars in Oshlack \etal (2002). The diagonal
line shows the results are identical. Lower panel: The  ratios of black hole masses 
estimated with two different relations are plotted against
 the radio-loudness of radio-loud quasars. The dashed line shows that two black hole 
masses are identical. Figure taken from Wu \etal (2004).}
\end{figure}

Brotherton (1996) studied the emission line properties of 59 radio-loud quasars. We 
adopted his published values of absolute V-band magnitude, equivalent width and FWHM of 
$H_{\beta}$ emission line. The
continuum luminosity at 5100$\AA$ and the $H_{\beta}$ luminosity 
were calculated after considering the Galactic extinction and K-correction 
(optical spectral index was 
assumed to be 0.3). We then estimated the BLR size using both $R-L_{H_\beta}$ and 
$R-L_{5100\AA}$ relations and derived the black hole mass with the formula
$M_{BH}=3V_{FWHM}^2 R/4G$ (here we assume the BLR velocity 
$V\sim (\sqrt{3}/2)V_{FWHM}$ as did by Kaspi \etal 2000). With these two relations, 
we also 
estimated the black hole masses of another 27 radio-loud quasars with available
data of both the equivalent width and FWHM of $H_{\beta}$ emission line in the Parkes 
Half-Jansky flat-spectrum Sample (PHFS) (Drinkwater \etal 1997; Francis, Whiting \& 
Webster 
2000; Oshlack \etal 2001). We compared the black hole masses obtained with the 
$R-L_{H_\beta}$ and 
$R-L_{5100\AA}$ relations in Figure 6. Evidently the masses obtained with
the  $R-L_{H_\beta}$ relation are systematically lower that those obtained with
the $R-L_{5100\AA}$ relation for  some extremely radio-loud quasars.
In Figure 6 we also show how the difference of black hole masses varies with
the radio-loudness for these two samples of radio-loud quasars.  It is clear that 
the difference of  black hole masses is smaller when the radio-loudness is small 
but becomes
larger as the radio-loudness increases. For some  individual quasars with 
higher radio-loudness, the black hole mass estimated with the  $R-L_{5100\AA}$ 
relation can be 3$\sim$10 times larger than that estimated with the 
$R-L_{H_\beta}$ relation.  
For radio-quiet AGNs,  however, both the optical continuum 
and emission line luminosities are probably free from the jet 
contributions and  therefore both can be good tracers of photo-ionization 
luminosity. Using these two relations will give almost identical results. 

More recently Kong \etal (2006) extended such a study to the broad UV emission
lines MgII and CIV. Their results confirm again that the relation between the BLR size
and broad emission line luminosity can be used to derive more accurate black hole mass
for extremely radio-loud AGNs. 

\section{Discussions}

We have presented some applications of using 
the empirical $M_{BH}-\sigma$ and $R-L$ relations
to estimate the black hole mass of various types of AGNs and reported some progress in
improving these studies by using the fundamental plane of the elliptical host galaxies
of AGNs and the relation between the BLR size and broad emission line luminosity. 
Although there are still some uncertainties,
these 
multiple methods can give consistent results on the black hole mass of most AGNs and have
been used to reliably estimate the black hole mass of  larger samples of AGNs.

\begin{figure}
\centering
\psfig{file=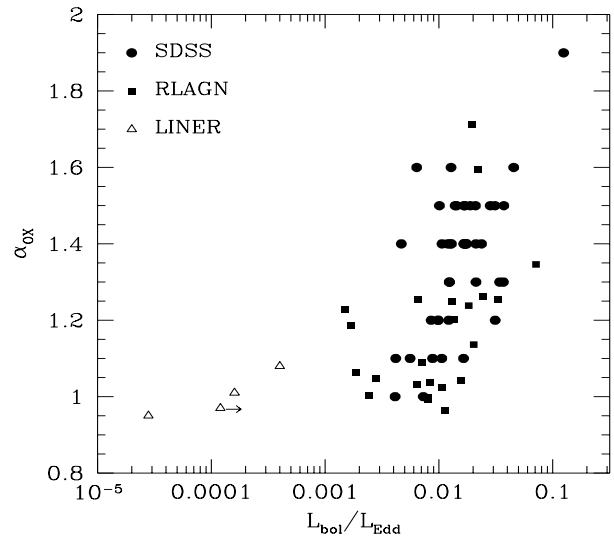,width=9.3cm,height=8cm,angle=0}
 \caption{Relation between the broad band spectral index 
$\alpha_{OX}$ and Eddington ratios for three 
different samples (SDSS, Radio-loud and LINER-type) of double-peaked broad line
AGNs (see section III). Figure taken from Wu \& Liu (2004).}
\end{figure}

Knowing black hole mass is important to study AGN physics. One of such examples is to
use the black hole mass to estimate the Eddington ratio ($L_{bol}/L_{Edd}$, which 
usually defines the dimensionless 
accretion rate of AGNs. From the accretion disk physics, we know that at different 
accretion rate, the accretion flow may have different accretion modes 
(Abramowicz \etal 1995).
Therefore, different Eddington ratios of AGNs may reflect different accretion physics.
Fig. 7 shows the dependence of the broad band spectral index $\alpha_{OX}$ on the
Eddington ratios for double peaked broad line AGNs with different luminosity. Clearly,
the AGNs at higher accretion rate have different spectral shape in the overall spectral
energy distributions (SED) from those at lower accretion rate. Wu \& Liu (2004) has 
demonstrated that the objects with higher $\alpha_{OX}$ values 
display big blue bumps in the optical/UV bands, which
is widely believed to be from the thermal radiation of an optically thick accretion 
disk (Shakura \& Sunyaev 1973).
On the contrary, the objects with smaller $\alpha_{OX}$ values (usually with smaller
Eddington ratios) may host an optically thin, radiatively inefficient 
advection dominated accretion flow (Narayan \& Yi 1994). Therefore, our knowledge of
the accretion rate based on the black hole mass estimates is crucial to understand
the accretion physics.

\begin{acknowledgments}
XBW thanks the organizers for kind invitation and Prof. 
Heon-Young Chang, Prof. Myeong-Gu Park and the 
staffs in APCTP for  hard works during the winter school. We are also 
grateful to Mr. Lei Qian and Mr. Bingxiao Xu for helpful discussions on black holes 
and accretion physics. This work is supported
by the NSFC Grants  (No. 10473001 \& No. 10525313), the RFDP
Grant (No. 20050001026) and the Key Grant Project of Chinese Ministry
of Education (No. 305001).
 
\end{acknowledgments}


\end{document}